\documentstyle[psfig]{article}
\newcommand{\beq}{\begin{equation}}
\newcommand{\eeq}{\end{equation}}
\newcommand{\beqa}{\begin{eqnarray}}
\newcommand{\eeqa}{\end{eqnarray}}

\begin{document}
\begin{center}
{\bf {\LARGE  AGAPE, an experiment to detect MACHO's in the direction of the
Andromeda galaxy}}
\\
{}~\\
{\large R. Ansari$^a$, M. Auri\`ere$^b$, P. Baillon$^c$, A. Bouquet$^d$,
G. Coupinot$^b$, \\C. Coutures$^e$, C. Ghesqui\`ere$^f$, Y.
Giraud-H\'eraud$^f$, P. Gondolo$^d$,
J. Hecquet$^b$,\\
J. Kaplan$^d$, A.L. Melchior$^d$, M. Moniez$^a$, J.P. Picat$^b$
and G. Soucail$^b$ }
\vspace{1cm}

$^a$ {\em LAL, Universit\'e Paris Sud, Orsay, France}, \\$^b$ {\em
Observatoire Midi-Pyr\'en\'ees Bagn\`eres de Bigorre et Toulouse,
France},\\$^c$ {\em CERN, Gen\`eve, Switzerland}, $^d$ {\em LPTHE,
Universit\'es Paris 6 et 7, France}, \\$^e$ {\em DAPNIA, CEN Saclay,
France}, $^f$ {\em LPC Coll\`ege de
France, Paris, France.}
\vspace{.5cm}

Presented at the 17th Texas Symposium on Relativistic Astrophysics, M\"unchen
germany, December 94, by J. Kaplan.
\end{center}
\vspace{.5cm}
PAR-LPTHE 95-07
\vspace{.5cm}

The M31 galaxy in Andromeda is the nearest large galaxy after the Small and
Large Magellanic
Clouds. It is a giant galaxy, roughly 2 times as large as our Milky Way, and
has its own halo.
As pointed by A. Crotts{\small\cite{crotts}} and independantly by some
of us{\small\cite{BBGK}} M31 provides a rich field of stars  to search for
MACHO's
in galactic halos by gravitational microlensing{\small\cite{pacz}}. M31 is a
target complementary to the Large Magellanic Cloud and the galactic
bulge wich are used by the three current
experiments{\small\cite{MACHO,EROS,OGLE}}.
It is complementary in that it allows to probe the halo of our galaxy
in a direction very different from that of the LMC. Moreover, the fact that M31
has
its own halo and is tilted with respect to the line of sight provides a
very interesting signature~: assuming an approximately spherical halo for M31,
the  far side
of the disk lies behind a larger amount of M31 dark matter, therefore more
microlensing
events are expected on the far side of the disk. Such an asymmetry could not be
faked by
variable stars{\small\cite{crotts}}.

In other words, M31 seems very appropriate to detect brown dwarfs through
microlensing.
However, as very few stars of M31 are resolved, we had to develop an approach
to look for
microlensing by monitoring the pixels of a CCD, rather than individual
stars{\small\cite{BBGK}}. The AGAPE collaboration has set out to implement
this idea.\\

{\bf \large MONITORING PIXELS}

In the case of a crowded field such as M31, the light flux $F_{\mbox{\rm
pixel}}$ on a pixel comes from  the
many stars in and around it, plus the sky background. The light
flux of an  individual
star, $F_{\mbox{\rm star}}$,  is spread among all pixels of the seeing spot and
only a
fraction of this light, $F_{\mbox{\rm pixel}} = \left\{  \mbox{\rm seeing
fraction}
\right\} \times \ F_{\mbox{\rm star}}$, reaches the central pixel.  If the star
luminosity is amplified by a factor $A$, the pixel flux
increases by :
\beq
\Delta F_{\mbox{\rm pixel}} = (A-1) \ \left\{ \mbox{\rm seeing fraction}
 \right\} \ F_{\mbox{\rm star}}
\eeq
The amplification of the star
luminosity allows an event to be detected if
the flux on the brightest pixel rises sufficiently high above its rms
fluctuation $\sigma_{\mbox{\rm pixel}}$ :
\beq
\Delta F_{\mbox{\rm pixel}} > Q\ \sigma_{\mbox{\rm pixel}} \label{detect}
\eeq
Typically, in our simulations, we require $Q$ to be larger than 3 during 3
consecutive
exposures and larger than 5 for at least one of them.\\

{\bf \large EXPECTED STATISTICS.}

We have performed numerical simulations using the above detection
criterium. As anticipated, the number of events we expect to be
able to detect depends strongly both on the stability of the pixel and on
the average seeing. The numbers in Table \ref{table2} below assume a two
meter telescope, a field of view of 60 by 20 arcminutes (planed for a
second generation experiment) centered on
the center of M31, 1 arcsecond pixels, 30 minutes exposures, and
120 consecutive nights. The brown dwarf mass is taken to b\-e 0.08 $M_\odot$,
and we assume a
standard halo (see reference{\small\cite{BBGK}} for instance) with a local dark
matter density of 0.3
GeV/cm$^3$ (0.0075 $M_\odot/pc^3$) and a core radius of 5 kpc.
\begin{table}[hbtp]
\caption{Expected number of events under various observing conditions
\label{table2}}
\begin{tabular*}{\textwidth}{@{}l@{\extracolsep{\fill}}rrrrr}
\hline
 seeing& $\sigma_{\mbox{\rm pixel}}/F_{\mbox{\rm pixel}}$ & number of
& number of & total  \\
& & galactic events & M31 events & \\
\hline
1'' & 2\%  &  9 & 15 & 24 \\
    & 1\%  & 15 & 29 & 44 \\
    & 0.5\%& 27 & 50 & 77 \\
\hline
2'' & 2\%  & 3  & 4  & 7  \\
    & 1\%  & 7  & 10 & 17 \\
    & 0.5\%& 11 & 21 & 32 \\
\hline
\end{tabular*}
\end{table}

These numbers have to be compared with the 2 events per year expected by  the
EROS
collaboration   for  brown dwarfs
with  masses of order 0.1 $M_\odot$.

It is clear from Table \ref{table2} that the number of detectable
events depends crucially on the relative flux fluctuation on the pixel,
$\sigma_{\mbox{\rm pixel}}/F_{\mbox{\rm pixel}}$. To study the feasibility of
the experiment, we have analyzed these pixel
fluctuations in three series of real data (Table \ref{table1}) :

$\bullet$ 1) 82 images of the Large Magellanic Cloud (LMC) taken by the EROS
collaboration

$\bullet$ 2) 26 images of M31 taken with the one meter telescope at Pic du
Midi, in
collaboration with F. Colas (Bureau des longitudes, Paris) and J. Lecacheux
(DESPA, Meudon)
\begin{table}[htb]
\caption{The mean relative fluctuation obtained for the three series of images
listed
above. The numbers in the first column refer to the numbers in the
list.\label{table1}}
\begin{tabular*}{\textwidth}{@{}l@{\extracolsep{\fill}}lccc}
\hline Images                & mirror size & pixel size        &relative
fluctuation\\
                      & (meter)     & (arcsec)          &                    \\
\hline LMC EROS ($\bullet$ 1)& 0.4         & 1.15              & 3\%
    \\ M31
Pic  ($\bullet$ 2)& 1           &  0.7              & 1\%                \\ M31
Pic
($\bullet$ 3) & 2           &  0.25             & 0.7\%              \\ M31 Pic
($\bullet$
3) & 2           &1 (``superpixels'')& 0.23\%             \\
\hline
\end{tabular*}
\end{table}

$\bullet$ 3) 4 images of M31 taken by E. Davoust (OMP, Toulouse) with the 2 m
telescope at
Pic du Midi. In this case, the angular size of the pixels is  very
small (0.23''), and we have also considered a
rearrangement in 4$\times$4 ``superpixels''.

The results given in Table \ref{table1} clearly show that the required
photometric stability of pixels  can be
reached. Moreover, the analysis of the third series of data shows
that pixels small compared to the seeing allow an efficient matching between
images,
whereas, once images are matched, superpixels are more  appropriate for a
stable photometry.\\

{\bf \large DISCRIMINATING AGAINST VARIABLE STARS AND OTHER VARIABILITIES}

Variable stars should be the main background. The usual tools to discriminate
against this
background are available (symmetry, unicity, achromaticity of the light curve).
Still,  some
points particular to our approach are  discussed below.

{\bf Achromaticity.} At first sight, one would think that there is no
achromaticity
as a star rising above a background of a different color will cause a color
variation of
the pixels involved. However, it is easy to show that when a star rises above
the
background in two color bands (say red and blue) then the ratio
\beq
\frac{\left(F_{\rm pixel} -\langle F_{\rm pixel} \rangle\right)_{\rm red}}
     {\left(F_{\rm pixel} -\langle F_{\rm pixel} \rangle\right)_{\rm
blue}} =
\frac{\left. {F_{\rm star}}\right|_{\rm red}}{\left. {F_{\rm star}}\right|_{\rm
blue}}
\eeq is constant in time during a microlensing event.

{\bf High amplifications.} To rise above the background an unresolved star
needs a
rather high amplification ($\langle A \rangle \sim 6$), which will exclude most
variable
stars.\\
\hbox{~~~} On the other hand small secondary maxima indicating unstable stars
with
occasional strong flares will not be discriminated. Further study is required
in this
respect. \\

{\bf \large THE FIRST RUN OF AGAPE}

We were given 57 half nights of observation on the 2 meter telescope ``Bernard
Lyot'' at
Observatoire du Pic du Midi in the French Pyr\'en\'ees, from September 29 to
November 24
1994. The field was $8' \times 8'$ only, covered by 4 exposures on a $800
\times 800$ thin Tektronix CCD camera with pixels  $0.3"$ wide.
\begin{figure}[htb]
\vspace{-0.7cm}
\begin{minipage}[t]{0.5\textwidth}
{\psfig{figure=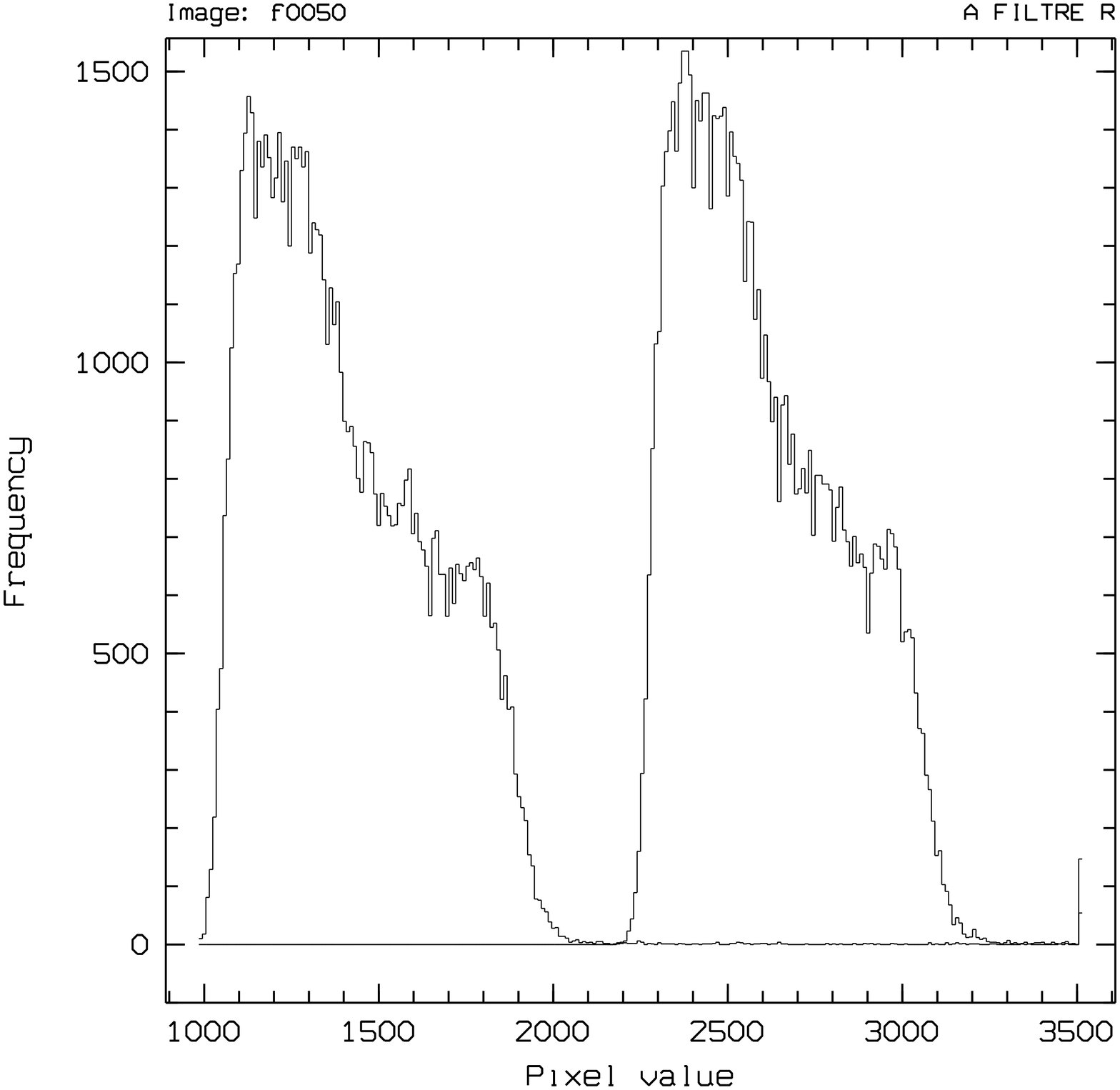,width=\textwidth}}
\end{minipage}\hfill
\begin{minipage}[t]{0.5\textwidth}
{\psfig{figure=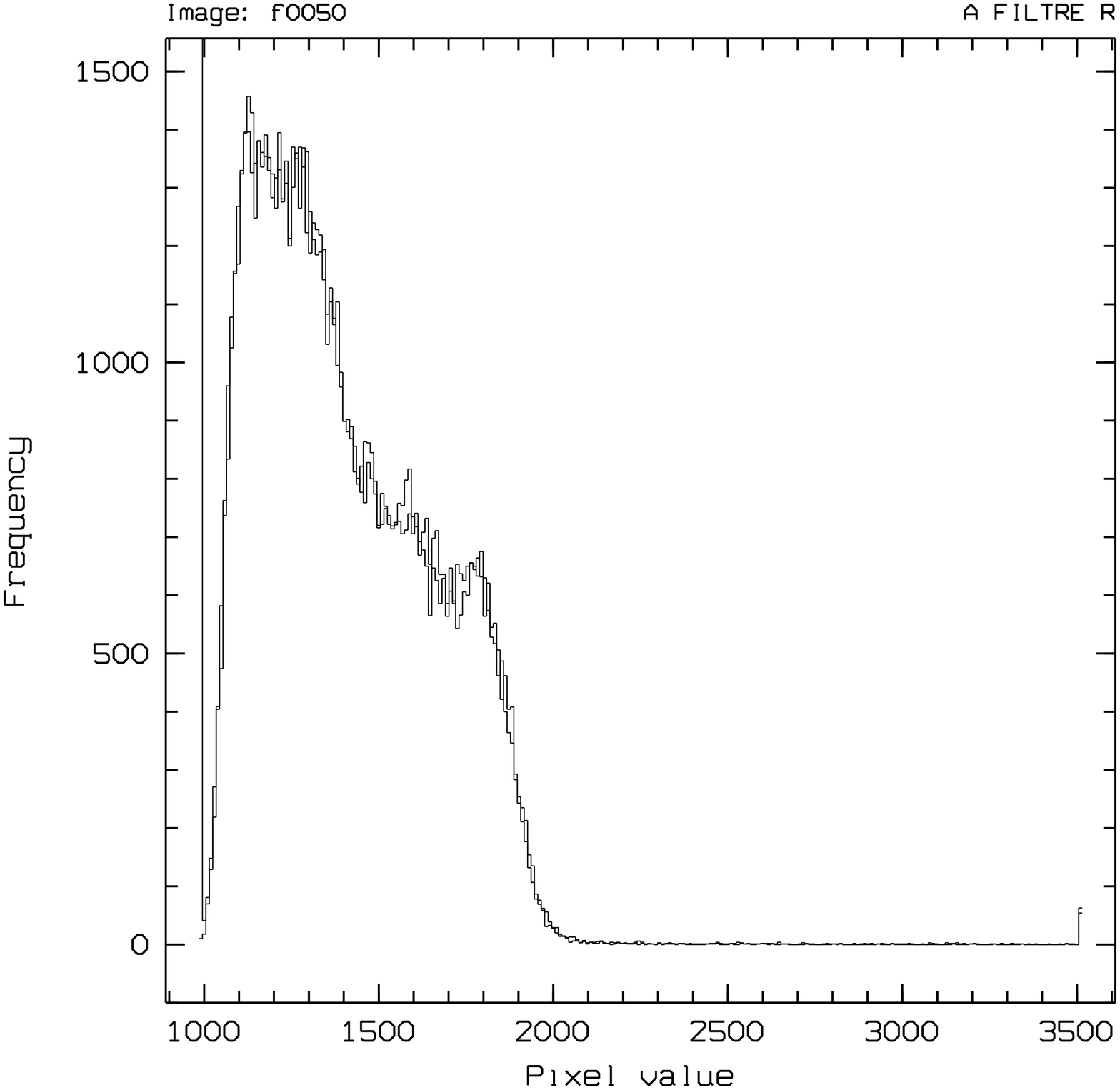,width=\textwidth}}
\end{minipage}\\
\hspace*{3.6cm} a \hspace*{6.6cm} b\\
\vspace{-6cm}
\hspace*{1.15cm}
\begin{minipage}{\textwidth}
{\psfig{figure=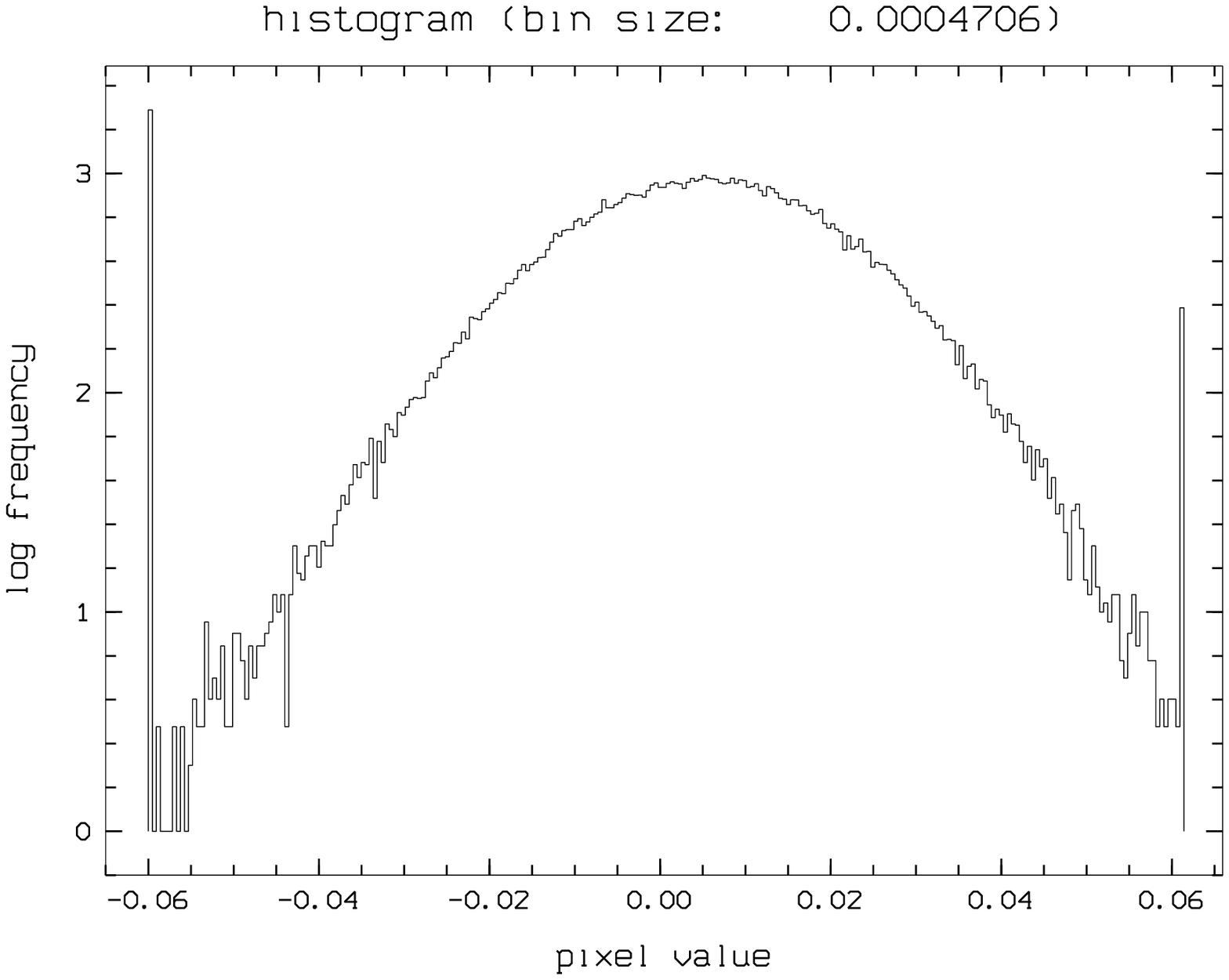,width=\textwidth}}
\vspace{0.3cm}
\hspace{3.5cm} c
\end{minipage}
\caption{photometric alignement of two different images of the same
field. Histogramms of pixel flux of two images, a: before treatment, b:
after photometric aligement. c: Histogram of the difference of pixel
fluxes between the two images, after photometric alignement. }\label{FIG1}
\end{figure}

The data of this prototype run are currently under treatment. A key
step of this treatment is the alignement of successive images both in
position and in photometry. The photometric alignement is performed by
linearly transforming the light flux of one image in such a way that
the mean flux and the variance of the transformed image matches those
of some reference image. The result of this alignement between images
is illustrated in figure 1. After alignement the dispersion of the
relative difference between the two images is 1.6\%. This
preliminary result is very encouraging as our alignement procedures are
not yet optimized.   \\

{\bf AKNOWLEDGMENTS} We thank The EROS collaboration, and E. Davoust who
allowed us to use their
data, as well as F. Colas, and J. Lecacheux with whom we took data on the 1
meter telescope at
Pic du Midi.  The help of F. Colas during our first observation run has been
particularly appreciated.


\begin{thebibliography}{99}
\vspace{-8pt}
\bibitem{crotts} Crotts A. P. S. 1992. ApJ. 399; L43.
\vspace{-8pt}
\bibitem{BBGK} Baillon P. Bouquet A. Giraud-H\'eraud Y. \& Kaplan J. 1993.
A\&A 277; 1.
\vspace{-8pt}
\bibitem{pacz} Paczy\'nski B. 1986. ApJ. 304; 1.
\vspace{-8pt}
\bibitem{MACHO} Alcock et al., 1993. Nature 365; 621, and these proceedings.
\vspace{-8pt}
\bibitem{EROS} Aubourg E. et al. 1993. Nature 365; 623.
\vspace{-8pt}
\bibitem{OGLE} Udalski A. et al. 1993. Acta Astronomica 43; 289.
\end{thebibliography}
\end{document}